# Simulation-based Verification of SystemC-based VPs at the ESL

Mehran Goli and Rolf Drechsler
University of Bremen/DFKI Bremen, Bremen, Germany

## Abstract

SystemC-based Virtual Prototypes (VPs) at the *Electronic System Level* (ESL) are increasingly adopted by the semiconductor industry. The main reason is that VPs are much earlier available, and their simulation is orders of magnitude faster in comparison to the hardware models at lower levels of abstraction (e.g. RTL). This leads designers to use VPs as reference models for early design verification. Hence, the correctness of VPs is of utmost importance as undetected errors may propagate to less abstract levels in the design process, increasing the fixing cost and effort. In this paper, we introduce a comprehensive simulation-based verification approach to automatically validate the simulation behavior of a given SystemC-based VP against both the TLM-2.0 rules and its specifications, i.e. functional and timing behavior of communications in the VP.

## 1  Introduction

Hardware modeling at the *Electronic System Level* (ESL) received strong attention in the last decades. In particular, modeling system as a *Virtual Prototype* (VP) in SystemC language using its *Transaction Level Modeling* (TLM) framework. The much earlier availability and the significantly faster simulation speed of VPs in comparison to the RTL hardware models are the main reasons that VPs are used as reference models for an early system verification in the design process. Hence, ensuring the correctness of VPs is of the utmost importance, as undetected faults may propagate to lower levels and become very costly to fix.

At the ESL, TLM-2.0 (as the current standard) provides designers with a set of standard interfaces and rules (TLM-2.0 base protocol) to model a VP based on abstract communication (i.e. transactions). This allows designers to abstract away the implementation details related to the computation of IPs and only focus on communication. Thus, communication (among different IP cores) is the main part of a VP model that must be verified. The first step to verify the communication in a given VP is to check whether or not they adhere to the TLM-2.0 rules. The TLM-2.0 standard comes with more than 150 rules that must be adhered to when a TLM model is implemented [1] and which define the expected behavior of a TLM-2.0 compliant model. Neither the SystemC compiler nor the TLM library detects TLM protocol violations that occur during execution. Manually verifying all rules and detecting the source of any given error is error-prone and expensive even for simple models and thus practically impossible for complex designs. Therefore, automated verification techniques that verify the compliance of a given ESL model with at least the base protocol are needed.

Moreover, as a VP is the first executable model of the design's specification – describing its functionality and timing behavior in terms of abstract communication – a functional assurance of the VP against its specifications is necessarily required, especially if the VP under development represents a safety-critical system. Therefore, to ensure the correctness of communication in a given VP, apart from validating the VP against TLM-2.0 rules (protocol validation), the functionality and timing behavior of the VP must be verified as well.

In general, the SystemC-based VP correctness can be ensured by two different approaches: formal verification and simulation-based verification (a process that called also validation). Formal approaches usually require to specify the model in formal semantics such as abstract state machines [2, 3] or IR models [4]. However, due to the object-oriented nature and event-driven simulation semantics, it is very challenging to verify a given SystemC VP formally.

In contrast, simulation-based verification approaches [5, 6, 7, 8, 9, 10, 13, 11] are still the predominant techniques to verify systems at the ESL as they scale very well with an arbitrary complexity of VPs. In simulation-based verification, the behavior of SystemC models is verified during the simulation. In this scope, assertion-based techniques [6, 7] are particularly well-suited for validation purposes. However, they come with some major drawbacks as the following. First, deriving assertions (i.e. properties) from TLM-2.0 rules or the design specifications usually requires manual effort by designers. Second, the generated assertions mostly need to be inserted manually to the VP. Third, in many cases, the SystemC kernel [5] or the SystemC library [7] needs to be modified to trace transactions accurately. This either relies on expensive manual processes or causes compatibility issues that overall reduce the degree of automation.

In this paper, we introduce a comprehensive simulation-based verification approach that automatically validates a given SystemC VP against the TLM-2.0 rules and its specifications. It consists of three main phases; data extraction and transformation, property generation, and validation. The focus of this work is to detect the errors related to the most common and essential fault types of communi-cation in a VP at the ESL; i.e dynamic rules (that cannot be checked statically e.g. during compilation time) related to the TLM-2.0 base protocol transactions (and its attributes), functionality and timing behavior. The approach is applied to several case studies, including a real-world VP.

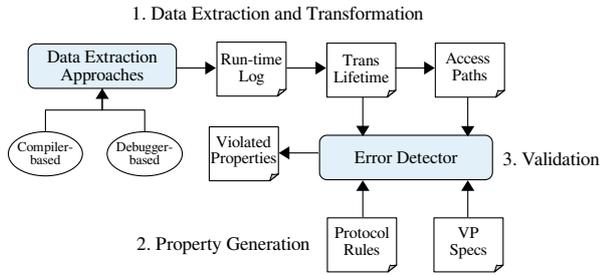

**Figure 1** Overview of the verification methodology.

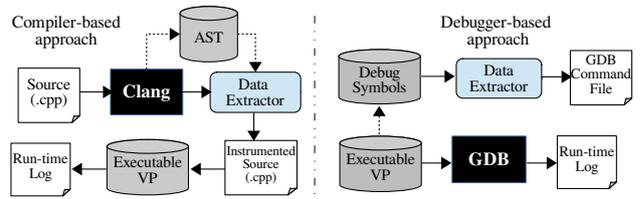

**Figure 2** Data extraction approaches.

## 2 SystemC and TLM-2.0

SystemC is a C++ based system level design language providing an event-driven simulation kernel. TLM-2.0 framework (as the current standard of SystemC TLM framework) introduces the transaction concept allowing designers to describe a model in terms of abstract communications using the base protocol and standard interfaces (e.g. *b_transport* and *nb_transport*). A transaction is a data structure (i.e. a C++ object) passed through TLM modules using function calls. A TLM module may include initiators (generating transactions), interconnects (acts as a transaction router), and targets (respond to the incoming transactions). Communication between two TLM modules in a VP can be performed based on two timing models, *Loosely-timed* (LT), and *Approximately-timed* (AT). The former is appropriate for the use case of software development while the latter for architectural exploration and performance analysis. While the LT model is implemented using the blocking transport, the AT model is implemented using the non-blocking transport providing multiple phases and timing points for a transaction. Due to the combination of these phases and timing points, 13 unique transaction types are defined in the base protocol.

## 3 Methodology

Fig. 1 provides an overview of the proposed approach includes three main phases which are 1) extracting the run-time behavior (i.e. transactions) of the SystemC VP and analyzing the extracted information to transform it into a set of intermediate representations, 2) Generating a set of properties from TLM-2.0 based protocol rules and VP specifications, and 3) Validating the VP's behavior against the generated properties.

### 3.1 Data Extraction and Transformation

The first phase of the proposed simulation-based approach is to access the run-time information of a given VP describing its behavior (which is defined in terms of transactions).

This requires tracing all transactions of the VP generated by different initiator modules and transferred through the interconnect to access the corresponding target modules. To trace a given VP's transactions, one of the following approaches can be used based on the designers requirements [12].

**Compiler-based approach:** Fig. 2 shows an overview of the Compiler-based approach consisting of two main steps. First, analyzing the *Abstract Syntax Tree* (AST) of a given VP to generate an instrumented version of the VP's source code. Next, compiling the instrumented source code with a standard C++ compiler (e.g., GCC or Clang) and executing it to log the run-time information. We refer the reader to [14, 15] for the details of the approach.

**Debugger-based approach:** As illustrated in Fig. 2, the static information of the compiled model is retrieved by analyzing its debug symbols to automatically generate a set of tailored GDB instructions. Next, the model is executed under the control of GDB using the previously generated instructions. The execution of the model is paused at certain events (such as function calls) to record the run-time information. We refer the reader to [16, 17] for the details of the approach.

In order to validate that a transaction adheres to the TLM protocol rules and the VP specifications, building the transaction lifetime is necessary. This requires analyzing the *Run-time Log* file to retrieve all information related to a transaction from the time that it is created by an initiator module until its completion. This time is considered as the transaction lifetime.

**Definition 1.** A transaction lifetime *TL* is a set of sequences *SQ* where
$$TL = \{SQ_i \mid 1 \leq i \leq n_T\}$$

and $n_T$ is defined based on which base protocol transaction is used as different types have disparate number of sequences. Since the validation of a given VP's transactions against the VP's specification requires to check whether or not the transactions are sent to the right target module (e.g. a right memory address) with the expected transaction type or delay w.r.t the VP specifications, we transform each transaction lifetime into an access path based on the following definition.

**Definition 2.** A complete simulation behavior of a given SystemC VP can be defined as a set of access paths *SAP* where each path *AP* shows a connection between an initiator module *IM* and a target module *TM* as below
$$SAP = \{AP_i \mid AP_i = \{IM \to TM :: (TID, TT, Tadrs, cmd, TD)\},$$
$$1 \leq i \leq n_{seq}\}$$

where *IM* and *TM* are initiator and target modules (their root and instance names), respectively. *TT* is the trans-action type illustrating which timing model (LT or AT) is used. To identify the transaction type, a unique type signature is generated by concatenating three parameters from the transaction lifetime i.e. *communication interface call*, *return status*, and *phase transitions*. The parameters *Tadrs*, *cmd*, *TD*, $n_{seq}$ indicate the address, transaction command, delay, and the number of sequence in a transaction lifetime, respectively.

## 3.2 Property Generation

The design rules are usually written in textbook specifications and designers use them to implement the design. To model a SystemC VP, a part of these specifications is defined by the TLM-2.0 base protocol describing e.g. how communications between TLM modules must be implemented. This type of constraint is implemented as a part of the *Error Detector* module (Fig. 1-phase 2) and explained in Section 3.3. The other parts of these specifications related to the functional and timing behavior of the VP are defined by designers and considered as *User constraints*.

The VP functional specifications $VP_{fs}$, for each initiator module *IM* include the list of all target modules *TM* that *IM* is allowed to access with a specific transaction type *TT* as below.

$$VP_{fs} = \{IM_i \mid IM_i \rightarrow \{(TM_j(address\_range), TT_n)\},$$
$$0 \leq i \leq n_{init}, 0 \leq j \leq n_{trg}, 0 \leq n \leq 13\} \quad (1)$$

Where $n_{init}$ and $n_{trg}$ indicate the number of initiator and target modules, respectively.

In order to validate the timing behavior of a given VP's transactions generated by different initiator modules against the VP's specifications, the timing specifications of the VP are required to be defined and given as inputs. This specification is defined in the same way as the functional specification. The only difference is that the required time of a communication between an initiator module and its corresponding target (total transaction delay) needs to be identified in the VP's specifications. Thus, the $VP_{ts}$ is defined as the following.

$$VP_{ts} = \{IM_i \mid IM_i \rightarrow \{(TM_j(address\_range), TT_n, TD)\},$$
$$0 \leq i \leq n_{init}, 0 \leq j \leq n_{trg}, 0 \leq n \leq 13\} \quad (2)$$

Where *TD* denotes the total delay of the generated transaction type *TT* by the initiator module *IM* to access the target module *TM*. We refer the reader to [9, 10] for the details of the approach.

## 3.3 Validation

The validation process of a given VP is performed in two main steps. First, the TLM-2.0 rules are checked indicating whether or not the VP behavior adheres to the TLM.2.0 based protocols. In this case, any violation is reported to designers to be overcome before the user constraints (defined based on the VP specifications) are verified. This type of constraint is directly generated from the TLM-2.0 base protocol including all rules related to the **transaction types** (e.g. the generated transaction of a given VP describes one of the valid based protocol transactions), **transaction attributes** (e.g. the data length attribute of a transaction must be a positive integer number) and the expected **TLM modules behavior** (e.g. an interconnect module must not modify the data attribute of a transaction). In the second step, user constraints are verified including both the functional and timing properties.

Constraints related to the transaction types are generated by translating the base protocol transactions into the corresponding type signature. This covers 25 rules of the TLM-2.0 based protocols. In order to check the correctness of each transaction lifetime against the transaction types fault, the following two steps are performed. First, the transaction type signature is generated by analyzing each transaction lifetime in *TL*. Then, an string matching algorithm is performed to identify unmatched transaction type signature that does not match the reference model. The lifetime of the violated transactions is analyzed to indicate the first faulty sequence. This sequence is reported to designers.

Verifying a VP against both functional and timing properties is performed by analyzing the access paths in *SAP*. For each property in *FP* or *TP* the *SAP* is traversed in order to find property violations related to the functional or timing behavior of the VP, respectively.

## 4 Experimental Results

To evaluate the quality of the proposed approach, we defined three type of faults FT1, FT2, and FT3 and injected them into the VPs. The verification methodology was used to validate the correctness of each VP against the TLM-2.0 rules and the VP's specifications. The definition of each fault model is as follows:

- FT1: an incorrect initialization of the transaction's response status (fault related to the transaction attributes rules), modification of the transaction data length by an interconnect module (fault related to the TLM modules behavior) and a wrong sequences order of transactions' phase transitions (fault related to the transaction type).

- FT2: initiating transactions with an incorrect address computation or an incorrect initialization of the VP memory configuration file.

- FT3: altering the timing annotation of transactions with an incorrect computation.

Experimental results for different types of ESL benchmarks are shown in Table 1. The first column shows three variants of SystemC VPs denoted as *FT1*, *FT2*, and *FT3* referring to the three faulty models, respectively. Columns *SystemC VP*, *Loc*, and *#Trans* list name, lines of code and the number of extracted transactions for each VP, respectively. The *#TT* column illustrates the number of transactions type implemented in each VP. Column *TM* presents the timing model of each design. Column *#Properties* shows the number of generated properties to validate each VP against the TLM-2.0 rules or its specifications. For this column, *Total*, *Pass*, *Fail* and *FTrans* illustrate the number of generated, satisfied, violated properties and faulty transactions, respectively. Overall, the total execution time

**Table 1** Experimental results of the proposed verification approach for all VPs

| Variant | VP Model | LoC | #Comp | #Trans | #TT | TM | #Properties Total | Pass | Fail | #FTrans | #Exe Time (s) CbA | DbA | CET |
|---|---|---|---|---|---|---|---|---|---|---|---|---|---|
| FT1 | Routing-model[1] | 456 | 6 | 10 | 1 | LT | 22 | 20 | 2 | 4 | 2.73 | 22.04 | 1.53 |
|  | AT-example[1] | 3,410 | 19 | 20 | 9 | AT | 35 | 32 | 3 | 7 | 23.02 | 473.23 | 17.19 |
|  | SoCRocket[2] | 50,000 | 20 | 200 | 8 | LT/AT | 55 | 52 | 3 | 37 | 63.15 | 5,498.41 | 27.80 |
| FT2 | Routing-model[1] | 456 | 6 | 10 | 1 | LT | 24 | 18 | 6 | 6 | 2.28 | 21.59 | 1.53 |
|  | AT-example[1] | 3,410 | 19 | 20 | 9 | AT | 84 | 69 | 15 | 8 | 19.25 | 469.46 | 17.19 |
|  | SoCRocket[2] | 50,000 | 20 | 200 | 8 | LT/AT | 612 | 536 | 76 | 42 | 55.02 | 5,491.28 | 27.80 |
| FT3 | Routing-model[1] | 456 | 6 | 10 | 1 | LT | 20 | 15 | 5 | 5 | 2.26 | 21.55 | 1.53 |
|  | AT-example[1] | 3,410 | 19 | 20 | 9 | AT | 40 | 30 | 10 | 10 | 19.20 | 469.41 | 17.19 |
|  | SoCRocket[2] | 50,000 | 20 | 200 | 8 | LT/AT | 370 | 311 | 59 | 59 | 54.95 | 5,491.21 | 27.80 |

1 and 2 provided by [18] and [19], respectively  **LoC:** Lines of Code  **#Trans:** Number of Transactions  **#TT:** Number of Transactions' Types  **TM:** Timing Model  **FTrans:** Number of Faulty Transactions  **CbA:** Compiler-based Approach
**DbA:** Debugger-based Approach  **CET:** Compilation and Execution Time without modification

using the compiler-based approach even for a complex VP (SoCRocket) is about a minute, providing a fast verification solution. In the case of debugger-based approach, the total execution time is still within a reasonable time frame. The debugger-based approach requires only the executable version of the VP, thus the original source code and workflow (e.g., SystemC library or compiler) stay untouched. The main problem with intrusive approaches that rely on altering e.g., the SystemC library, interfaces, simulation kernel, or compiler is that these modifications may cause an issue for the application of several approaches in parallel, future updates or restrictive environments. Moreover, they mostly reduce the degree of automation as they require manual effort by designers. In the case of third-party IPs or legacy models where the source code may not be available at all, this approach is the only applicable solution. On the other hand, the compiler-based approach is very fast and scales well with an arbitrary complexity of VPs. However, it requires the availability of the VP's original source code. Since the proposed approach modifies neither the SystemC library nor the SystemC simulation kernel nor compiler, any results obtained using the approach are identical to the reference results in terms of VP's timing behavior and its functionality. Overall, due to the designer's concerns and requirements, they have the option to choose either the debugger-based or the compiler-based approach. All the experiments have been carried out on a PC equipped with 8 GB RAM and an Intel core i7 CPU running at 2.4 GHz.

## 5 Conclusion

In this paper, a comprehensive verification approach was presented, enabling designers to validate a given SystemC-based VP implemented using SystemC TLM-2.0 framework against the TLM-2.0 rules, and the VP's specifications. We showed how the simulation behavior of VPs based on transactions can be extracted using the debugger- and compiler-based approaches. The extracted information is translated into a set of transactions' lifetime and access paths to be validated against the TLM-2.0 rules and the VP's specifications, respectively. We demonstrated the effectiveness and scalability of our approach on several standard VPs including a real-world system.

## Acknowledgment

This work was supported in part by the *German Federal Ministry of Education and Research* (BMBF) within the project VerSys under contract no. 01IW19001 and by the University of Bremen's graduate school *System Design* (SyDe).